\documentstyle[sprocl]{article}

\input{epsf}
\begin{document}

\title{LOCALIZATION OF ELECTRONIC WAVE FUNCTIONS\\ ON QUASIPERIODIC LATTICES}

\author{\sc Thomas Rieth, Uwe Grimm, Michael Schreiber}

\address{Institut f\"ur Physik, Technische Universit\"{a}t, 
D-09107 Chemnitz, Germany}

\maketitle\abstracts{We study electronic eigenstates on quasiperiodic
lattices using a tight-binding Hamiltonian in the vertex model.  In
particular, the two-dimensional Penrose tiling and the
three-dimensional icosahedral Ammann-Kramer tiling are considered.
Our main interest concerns the decay form and the self-similarity of
the electronic wave functions, which we compute numerically for
periodic approximants of the perfect quasiperiodic structure. In order
to investigate the suggested power-law localization of states, we
calculate their participation numbers and structural entropy. We also
perform a multifractal analysis of the eigenstates by standard
box-counting methods. Our results indicate a rather different
behaviour of the two- and the three-dimensional systems. Whereas the
eigenstates on the Penrose tiling typically show power-law
localization, this was not observed for the icosahedral tiling.}

\section{Introduction}

There are, at least, three popular, yet apparently unrelated,
approaches aimed at a better understanding of the peculiar transport
properties of quasicrystals. Among those is the Hume-Rothery picture,
which gives a plausible explanation for the presence of the pseudo-gap
in the electronic density of states at the Fermi level. Another
approach considers quasicrystals as a conglomerate or as an
hierarchical arrangement of clusters.  Here, we follow a third
approach by investigating localization effects of electronic states
caused by quasiperiodicity. For this purpose, we consider
tight-binding Hamiltonians on periodic approximants of the
two-dimensional Penrose and the icosahedral Ammann-Kramer-Neri (AKN)
tiling and perform a statistical analysis of their eigenstates.

\section{Tight-binding model}

In the vertex model, an atomic orbital $|j\rangle$ is placed at every
vertex $j$ of the quasiperiodic tiling, and the Hamiltonian has the form
\begin{equation}
{\cal H}\; =\; 
\sum_{j,k=1}^{N}\, |j\rangle\, t_{jk}^{}\, \langle k|
\; +\; \sum_{j=1}^{N}\, |j\rangle\, V_j^{}\, \langle j|
\label{H} 
\end{equation}
where the hopping amplitudes are $t_{jk}^{}=1$ for vertices $j$ and $k$
connected by a bond, and $t_{jk}^{}=0$ otherwise. The on-site potentials
are chosen as $V_j^{}=0$, hence the spectrum
of ${\cal H}$ is symmetric around $E=0$. 
A simple characterization of an eigenstate $|\psi\rangle$ of ${\cal H}$ 
is given by the {\it participation number\/} $P$, defined by
\begin{equation}
P^{\, -1} \; = \; \sum_{j=1}^{N} \, \psi_j^{\, 4}\, ,
\qquad
 |\psi\rangle\; =\;\sum_{j=1}^{N}\, \psi_j^{}\, |j\rangle\, ,
\label{P}
\end{equation}
which can be interpreted as an estimate of the number of vertices
which carry a significant part of the wave function amplitude.
With increasing system size $N$, the participation number grows linearly
with $N$ for {\it extended\/} states, while it approaches a constant
for {\it localized\/} states.

Another interesting quantity is the so-called 
{\it structural entropy\/}\thinspace\thinspace\cite{RS3}
\begin{equation}
S_{\mbox{\scriptsize str}} \; = \; 
-\sum_{j=1}^{N}\, |\psi_j|^2\, \ln |\psi_j|^2
\;-\; \ln P \, ,
\label{str}
\end{equation}
the first term being the familiar Shannon entropy.
However, while plots of $S_{\mbox{\scriptsize str}}$ against the
{\it participation ratio\/} $p=P/N$ may look different for various
decay forms of the wave functions, this need not be the case and 
one has to be very careful if one intends to extract 
conclusive information on the decay form.

In the box-counting method, the system is divided into boxes of
linear size $\delta$. We denote the probability amplitude in the $k$\/th box
by $\mu_k(\delta)$. Its normalized $q$\/th moment $\mu_k(q,\delta)$
constitutes a measure. {}From this, one obtains the 
Lipshitz-H\"{o}lder exponent or 
singularity strength $\alpha$ of an eigenstate and the corresponding
fractal dimension $f$ by
\begin{eqnarray}
\alpha(q) & = & \lim_{\delta\rightarrow 0}\:
\textstyle \sum_k\, \mu_k(q,\delta)\, \ln\mu_k(1,\delta)/\ln\delta\, ,
\\
f(q)      & = & \lim_{\delta\rightarrow 0}\:
\textstyle \sum_k\, \mu_k(q,\delta)\, \ln\mu_k(q,\delta)/\ln\delta\, ,
\end{eqnarray}
yielding the characteristic singularity
spectrum $f(\alpha)$ in a parametric representation.

\section{Results: Penrose and Ammann-Kramer-Neri tiling}

For the Penrose tiling, roughly 10\% of the spectrum consists of
degenerate states in the band center ($E=0$),\thinspace\cite{ATFK88} 
separated from the rest of the band by a finite energy gap. The
degenerate eigenstates are combinations of strictly localized (confined) 
states.\thinspace\cite{ATFK88,RS1} 
For the AKN tiling, one also observes a small
number of degenerate states in the band center, but there is no indication 
of an energy gap in the spectrum.\thinspace\cite{RS3} 

In Ref.~4, the scaling behaviour of the participation number
was discussed for the two examples. For the Penrose
case, a power-law behaviour $P\sim N^{\beta}$ was found, with an
exponent $\beta$ that is smaller than one, decreasing towards
the band center. This indicates that the envelope of the 
eigenfunction shows a power-law decay. A closer investigation 
reveals that the state at the band edge is in fact 
extended.\thinspace\cite{RS3} 
The situation is very different for the AKN tiling: here the participation
number of states close to the band edge scales with $\beta<1$,
hinting at a power-law decay, while $\beta$ quickly grows to 
$\beta\approx 1$ as one moves towards the band center, compatible with
the presence of  extended states.\thinspace\cite{RS3,RSA}

\begin{figure}[tb]
\centerline{\epsfxsize=0.5\textwidth 
\epsfbox[50 84 470 407]{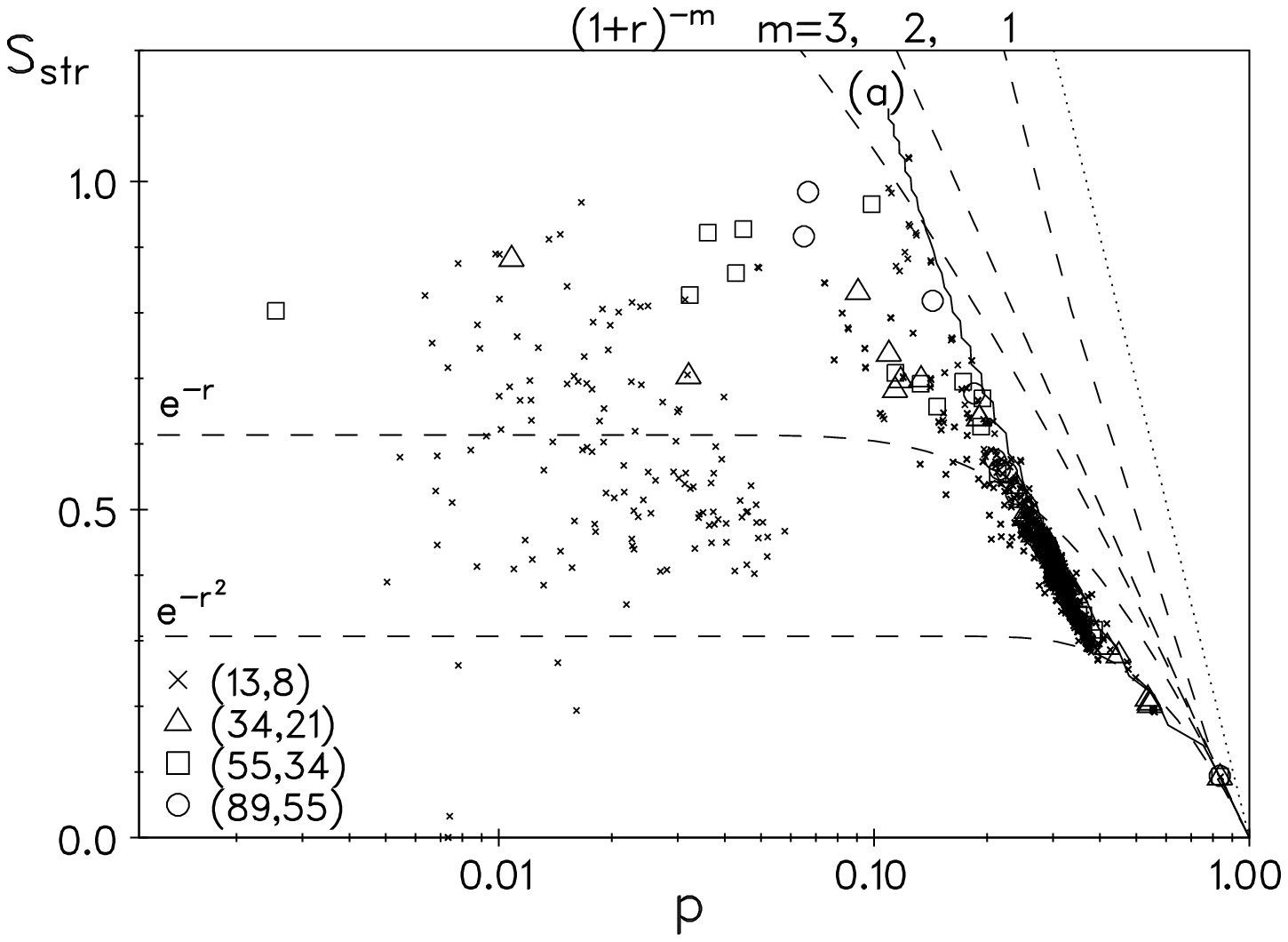}\hspace*{\fill}
\epsfxsize=0.5\textwidth\epsfbox[50 84 470 407]{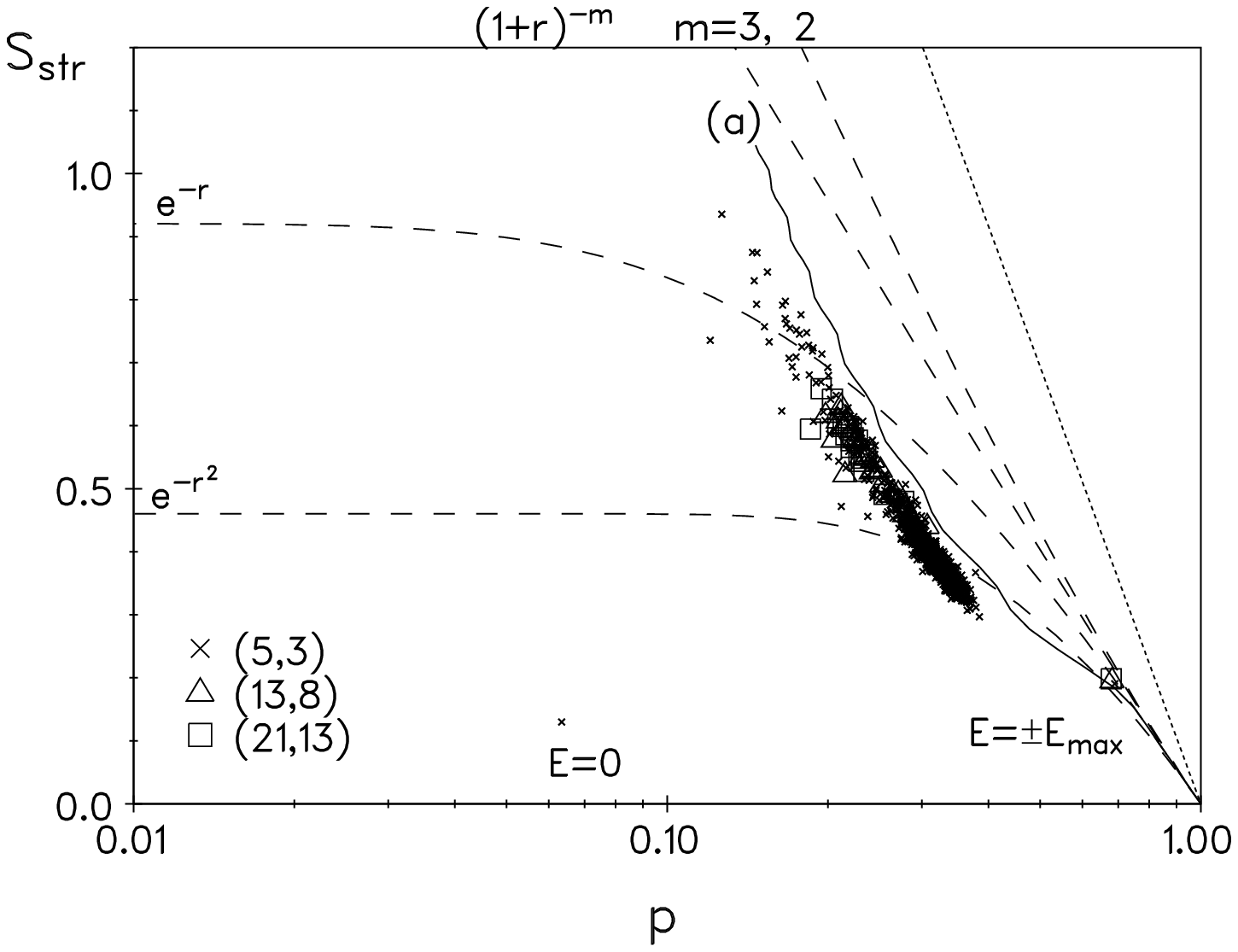}}\vspace*{-2ex}
\caption{Structural entropy $S_{\mbox{\scriptsize str}}$
[Eq.~(\protect\ref{str})] as function of the
participation ratio $p=P/N$ [Eq.~(\protect\ref{P})] for periodic
approximants of the  Penrose (left) and the AKN (right) 
tiling, compared with the dependence (dashed lines) of wave functions
of different shapes.\hspace*{\fill}\label{fig1}}
\end{figure}

Fig.~\ref{fig1} clearly shows a strong correlation between
the structural entropy $S_{\mbox{\scriptsize str}}$ [Eq.~(\ref{str})] 
and the participation ratio $p=P/N$. For both examples, this behaviour
can fairly well be reproduced assuming a decay form
\begin{equation}
|\psi(r)|^2 \;=\;
\left\{ \begin{array}{@{}l@{\quad}l} 
\exp(-r) & \mbox{for $r<R$}\\
C\left[\cos(cr+\phi)+1\right]r^{-2\alpha} & \mbox{for $r\ge R$}
\end{array}\right.
\end{equation}
with $R=0.75$, $c=10$, $\alpha=0.65$ for the Penrose case, and
$R=2.5$, $c=25$, $\alpha=0.8$ for the AKN tiling, corresponding
to the curves labeled ($a$) in Fig.~\ref{fig1}.

\begin{figure}[tb]
\centerline{
\begin{minipage}{0.55\textwidth}
\centerline{\epsfxsize=\textwidth\epsfbox{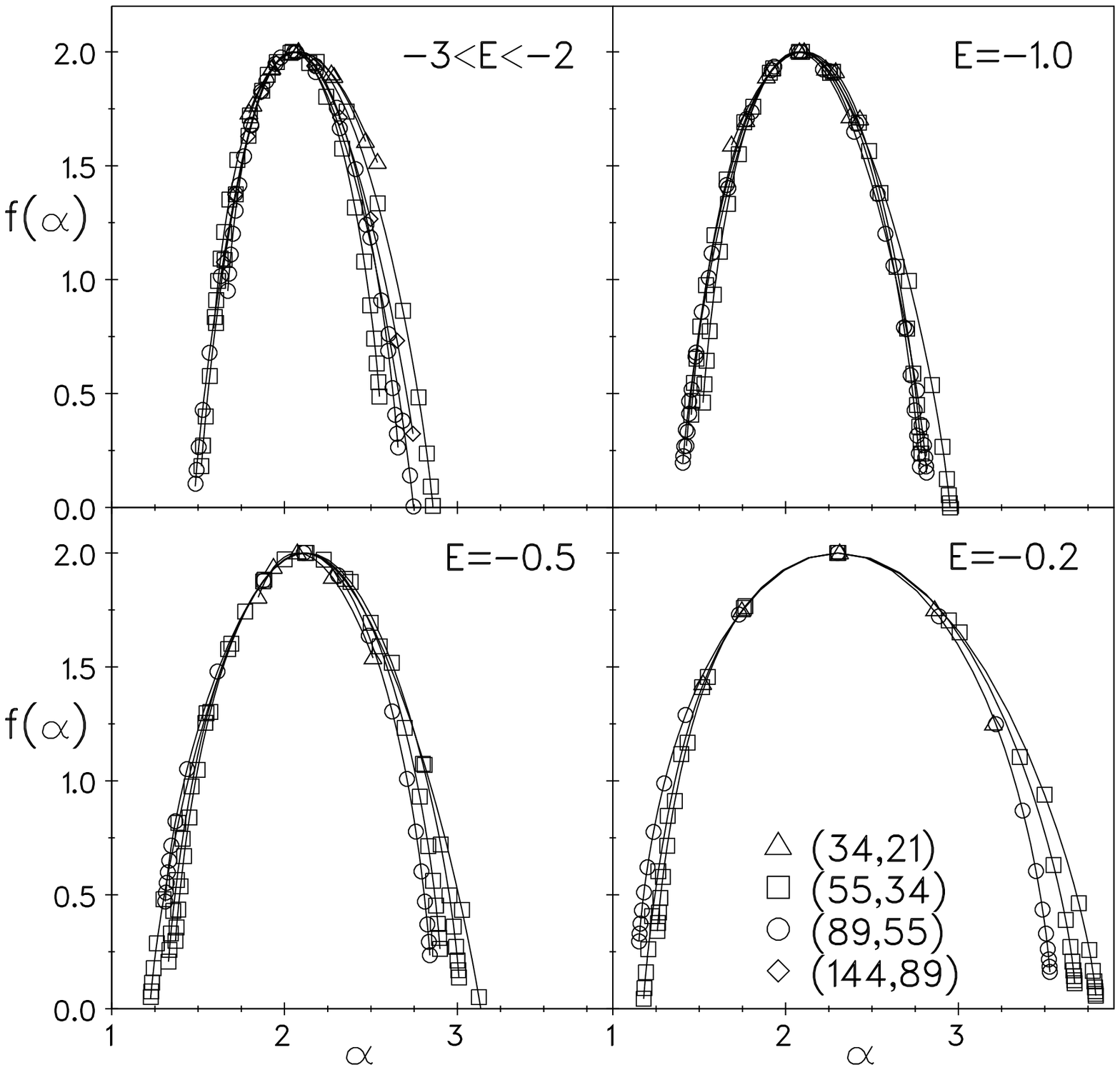}}
\end{minipage}\begin{minipage}{0.45\textwidth}
\centerline{\epsfxsize=\textwidth\epsfbox{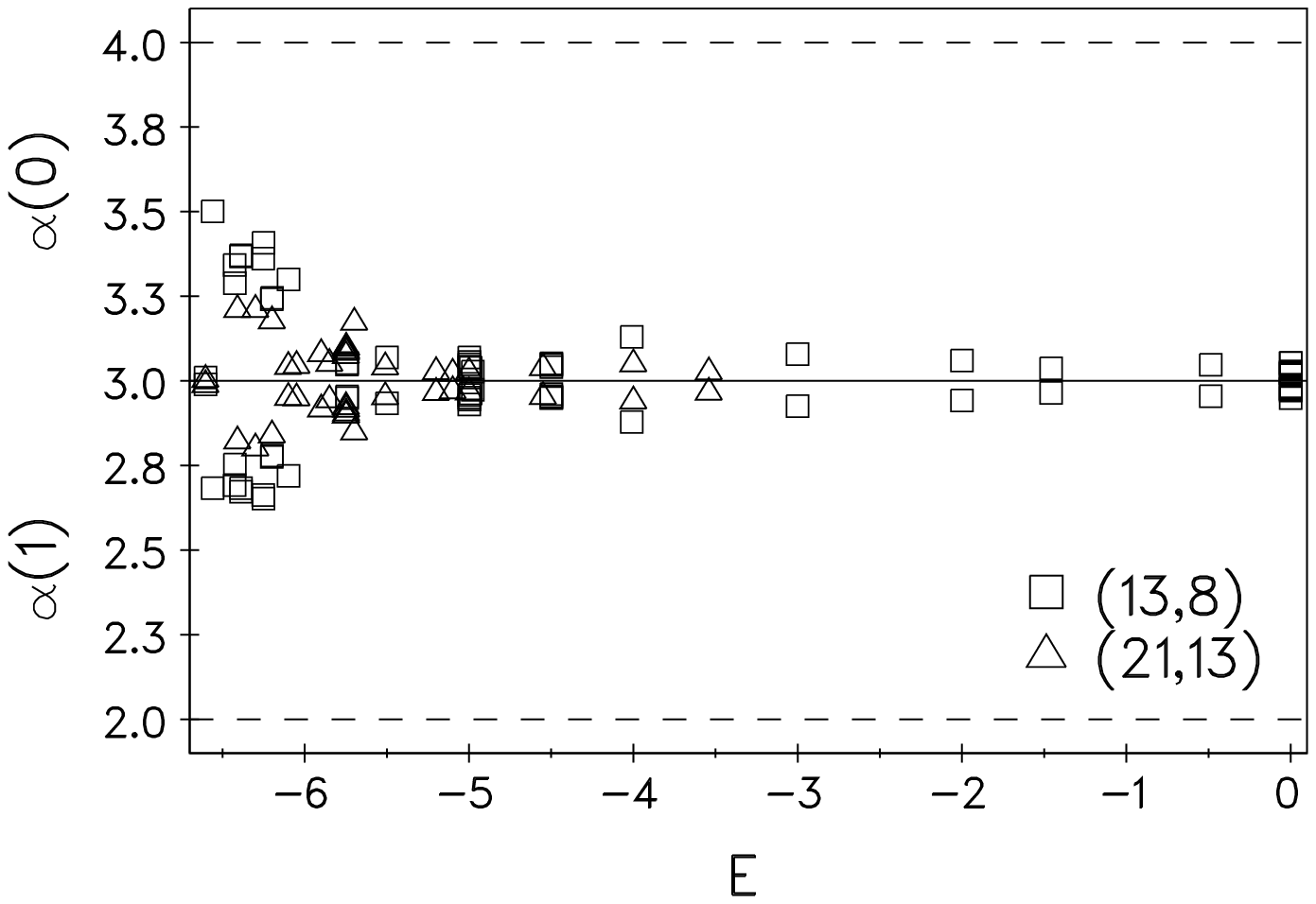}}
\hspace*{0.1\textwidth}\begin{minipage}{0.9\textwidth}
\caption{Singularity spectra $f(\alpha)$ and singularity strengths
$\alpha(0)$ and $\alpha(1)$ for various eigenstates of
periodic approximants of the Penrose (left) and  the AKN (right) tiling,
respectively.\hspace*{\fill}\label{fig2}}\vspace*{1.8ex}\end{minipage}
\end{minipage}}\vspace*{-0.4ex}
\end{figure}

Fig.~2 shows the results of a box-counting 
multifractal analysis on the eigenstates. The singularity spectra obtained
for the Penrose case are almost independent of the size of the systems,
and become wider as one moves from the edge to the center of the band.
In contrast, the states in the AKN tiling, apart from those
close to the band edge, show the behaviour of extended states; the
singularity strengths $\alpha(0)$ and $\alpha(1)$ shown in Fig.~2 
are very close to $3$, hence we cannot conclude a multifractal structure
of the states.

\section{Concluding remarks}

Typical eigenstates of a vertex-type tight-binding model on the
Penrose tiling have multifractal properties and show a power-law 
localization. This is in accordance with results on analytically
constructed eigenstates.\thinspace\cite{TFA88}

{}From our numerical analysis, we cannot draw similar conclusions
for the AKN case. While the scaling behaviour of the participation number
and the multifractal analysis favour extended states, the structural
entropy hints at a power-law decay. In order to answer this question, 
other quantities have to be investigated which are sensitive also to
weak power-law decays. Introducing disorder via the on-site 
term $V_j^{}$ leads to a localization transition which
appears to be very similar to the 
metal-insulator transition observed in a simple cubic 
lattice.\thinspace\cite{RS2} 
This corroborates the notion that the eigenstates in the AKN tiling without 
disorder are extended.

\end{document}